\documentclass[journal=jacsat,manuscript=article]{achemso}

\usepackage{lineno}
\usepackage{amsmath}
\usepackage{amssymb}
\usepackage{setspace}
\usepackage{amssymb} 
\usepackage{tabularx} 
\usepackage{soul,color}
\usepackage{float} 
\usepackage{multicol}
\usepackage{subcaption}
\usepackage{stfloats}
\usepackage{xcolor} 

\usepackage{hyperref}
\hypersetup{
	colorlinks   = true,  
	urlcolor     = blue,  
	linkcolor    = blue, 
	citecolor    = blue    
}

\usepackage[utf8]{inputenc}
\DeclareUnicodeCharacter{2005}{} 
\DeclareUnicodeCharacter{2212}{}

\usepackage{graphicx}
\graphicspath{{./Images/}}

\usepackage[version=3]{mhchem} 

\author{M.D. Hashan C. Peiris}
\affiliation{Materials Science and Engineering, Binghamton University, NY}

\author{Michael Woodcox}
\affiliation{Materials Measurement Laboratory, National Institute of Standards and Technology, Gaithersburg, MD}

\author{Diana Liepinya}
\affiliation{Materials Science and Engineering, University of Maryland, MD}

\author{Robert Shepard}
\affiliation{Department of Physics, Binghamton University, NY}

\author{Hao Liu}
\affiliation{Department of Chemistry, Binghamton University, NY}
\alsoaffiliation{Materials Science and Engineering, Binghamton University, NY}

\author{Manuel Smeu}
\email{msmeu@binghamton.edu}
\affiliation{Department of Physics, Binghamton University, NY}
\alsoaffiliation{Materials Science and Engineering, Binghamton University, NY}

\title[An \textsf{achemso} demo]
  {Aluminum oxide coatings on Co-rich cathodes and interactions with organic electrolyte}
\abbreviations{XPS, DFT, AIMD, EC, DMC, LNO, LCO, CO}


\begin{document}

\begin{tocentry}
Some journals require a graphical entry for the Table of Contents. This should be laid out ``print ready'' so that the sizing of the text is correct.
Inside the \texttt{tocentry} environment, the font used is Helvetica 8\,pt, as required by \emph{Journal of the American Chemical Society}.
The surrounding frame is 9\,cm by 3.5\,cm, which is the maximum permitted for  \emph{Journal of the American Chemical Society} graphical table of content entries. The box will not resize if the content is too big: instead it will overflow the edge of the box.
This box and the associated title will always be printed on a separate page at the end of the document.
\end{tocentry}
\newpage
\singlespacing
\begin{abstract} 
Lithium-ion batteries (LIBs) have become essential in modern energy storage, yet their performance is often constrained by the stability and efficiency of their components, particularly the cathode and electrolyte. Transition metal layered oxide cathodes, a popular choice for LIBs, suffer from several degradation mechanisms, including capacity fading, reactions with the electrolyte, unstable cathode electrolyte interfaces, and lattice breakdown during cycling. Additionally, the commonly used organic electrolytes in LIBs are prone to side reactions, especially at higher states of charge, further hindering performance. In recent years, oxide coating, such as alumina, has emerged as a promising strategy to enhance the durability of cathodes by forming a protective layer that mitigates detrimental reactions and improves the stability of the cathode electrolyte interphase (CEI). This study employs \textit{ab initio} molecular dynamics (AIMD) simulations to investigate the chemical and mechanical behavior of LiCoO$_2$ cathodes with and without aluminum oxide coatings in contact with an organic electrolyte. We examine the interactions between electrolyte molecules with both bare and coated cathode surfaces, focusing on the decomposition of ethylene carbonate (EC) and dimethyl carbonate (DMC), the formation of oxygen species, and solvation dynamics. We evaluate the mechanical robustness of the cathode-coating interface using axial strain and cleavage energy calculations. Our findings reveal that alumina coatings effectively reduce electrolyte degradation and stabilize the cathode structure, particularly under high-charge states. The coating's thickness and structural orientation are critical in enhancing the mechanical strength and minimizing the detrimental reactions at the cathode-electrolyte interface. These insights contribute to developing more durable LIBs by optimizing the interface chemistry and mechanical properties, providing a pathway toward higher energy densities and longer cycle life.
\end{abstract}

\newpage
\section{Introduction}
\par Ever since being commercialized in 1990, Li-ion batteries (LIBs) have been undergoing rapid research into optimizing their performance in every sense of the word\cite{west_electrochemical_2011,kaur_reviewsurface_2022}. Even though major improvements have been made, the traditional limitations of battery performance remain to be fully solved. Among these critical issues is the performance of the battery's cathode. Specifically, for transition metal (TM) layered oxide cathodes, several issues, such as fading capacity, reactions with the electrolyte molecules, unsustainable cathode-electrolyte interphase, cathode cracking, and lattice breakdown during cycling, have been reported to be particularly detrimental to the performance \cite{e_spotte-smith_critical_2024,xu_understanding_2016,sharifi-asl_facet-dependent_2017}. 

\par In addition to the role played by the cathode, the electrolyte used in a LIB has significant implications for the performance of the battery. While organic electrolytes have enabled better performance in commercialized LIBs, they still exhibit degenerative behavior, such as undesirable side reactions with the cathode \cite{freiberg_singlet_2018,jung_temperature_2018,rinkel_two_2022}. This gets exacerbated at higher states of charge (SOC) and is a critical issue in traditional and novel cathodes. These cathodes can be charged at higher potentials but are limited from doing so by the reactivity of the electrolyte at higher SOC. 

\par In search of performance improvements, in recent years, the application of coatings on the cathode surface has been suggested to improve battery cycling. Oxide coatings, in particular, have been proposed to result in remarkable performance improvements \cite{ge_advances_2023,johnson_brief_2014}. These are thought to act as an artificial cathode-electrolyte interphase (CEI) layer that exhibits favorable properties similar to those of typical CEI but mitigates its harmful aspects.  While its protective characteristics have been widely observed and reported, atomistic insights into its functioning have not been widely reported. 

\par Coating layers, including oxides/fluorides and lithium salts, serve as barriers that prevent direct contact between the cathode and the electrolyte or as conductive layers that improve ionic and electrical conductivity\cite{kaur_reviewsurface_2022}. Alumina interacts with LiOH and Li$_2$CO$_3$ within nickel-rich cathode materials under high-temperature conditions. Known for its hardness and resistance to chemical attacks from acidic and alkaline substances, alumina coatings improve the chemical stability of interfaces while promoting lithium-ion (Li$^+$) diffusion paths. Research has previously shown that applying a 1.5 wt.\% Al$_2$O$_3$ coating on over-lithiated Li$_{1.17}$Ni$_{0.135}$Mn$_{0.56}$Co$_{0.135}$O$_{2}$ cathode material enhances Li$^+$ diffusion and increases the specific exchange current at lower temperatures relative to uncoated cathode materials \cite{kaur_reviewsurface_2022,west_electrochemical_2011}. Ultra-thin coatings have been demonstrated to be the most effective in lithium nickel manganese cobalt oxide 
(NMC) cathodes, balancing the diffusion of Li through the coating while protecting the cathode surface from attack by the electrolytes.


\par In this work, we present a comprehensive investigation into the chemical and mechanical dynamics of the cathode electrolyte interface and the effects of the aluminum oxide coating on the cathode. First, we present the chemical interactions on the interfaces observed in our trajectories with respect to the electrolyte. We discuss the breakdown of ethylene carbonate (EC) and dimethyl carbonate (DMC) molecules on the exposed cathode and coated surface. Thereafter, we discuss the breakdown of the exposed cathode surface in the trajectory and how it contrasts with the coated cathode surface. In particular, we discuss the formation of molecular oxygen on the surfaces and the source of emissions. We use charge analysis schemes to identify the charge transfer mechanisms of such processes. We then analyze the formation of solvation shells, their persistence with time, and their behavior at the interface. Secondly, we present our findings on the mechanistic properties of the cathode and the oxide coating using mechanically controlled break junction simulations to elucidate the bonding properties of the coating to the cathode and the effects of lithiation. 

\section{Methods}
 

\par The organic electrolyte was modeled using the Vienna Ab initio Simulation Package (VASP)\cite{Kresse1994,Kresse1996a,Kresse1996b}. Projector augmented wave (PAW) potentials were used to mimic the ionic cores, and the Perdew-Burke-Ernzerhof (PBE) generalized gradient approximation (GGA) provided the exchange and correlation functional for all structural relaxations and energy calculations \cite{Hafner1994,Kresse1999,PBE1,PBE2}. A $\Gamma$-centered 1$\times$1$\times$1 Monkhorst-Pack \textit{k}-point grid was used for all calculations \cite{Pack1977,Pack1976}. The Brillouin zone was sampled using Gaussian smearing.

\par All structural relaxations and energy calculations were performed using a plane wave energy cutoff set at 700~eV. The primitive cells for the cathodes were obtained from the Materials Project \cite{Jain2013}. To account for the highly correlated electronic states present in the layered transition metal oxide systems such as LiCoO$_2$, a Hubbard \textit{U} value of 3.2~eV was used for the AIMD trajectories. Setup, visualization, and manipulation of the systems were done using Ovito\cite{Stukowski2009} and Vesta\cite{Momma2011}, with Avogadro\cite{Hanwell2012} used to create the electrolyte configuration.

\begin{figure}[htbp]
    \centering
    \includegraphics[scale=0.6]{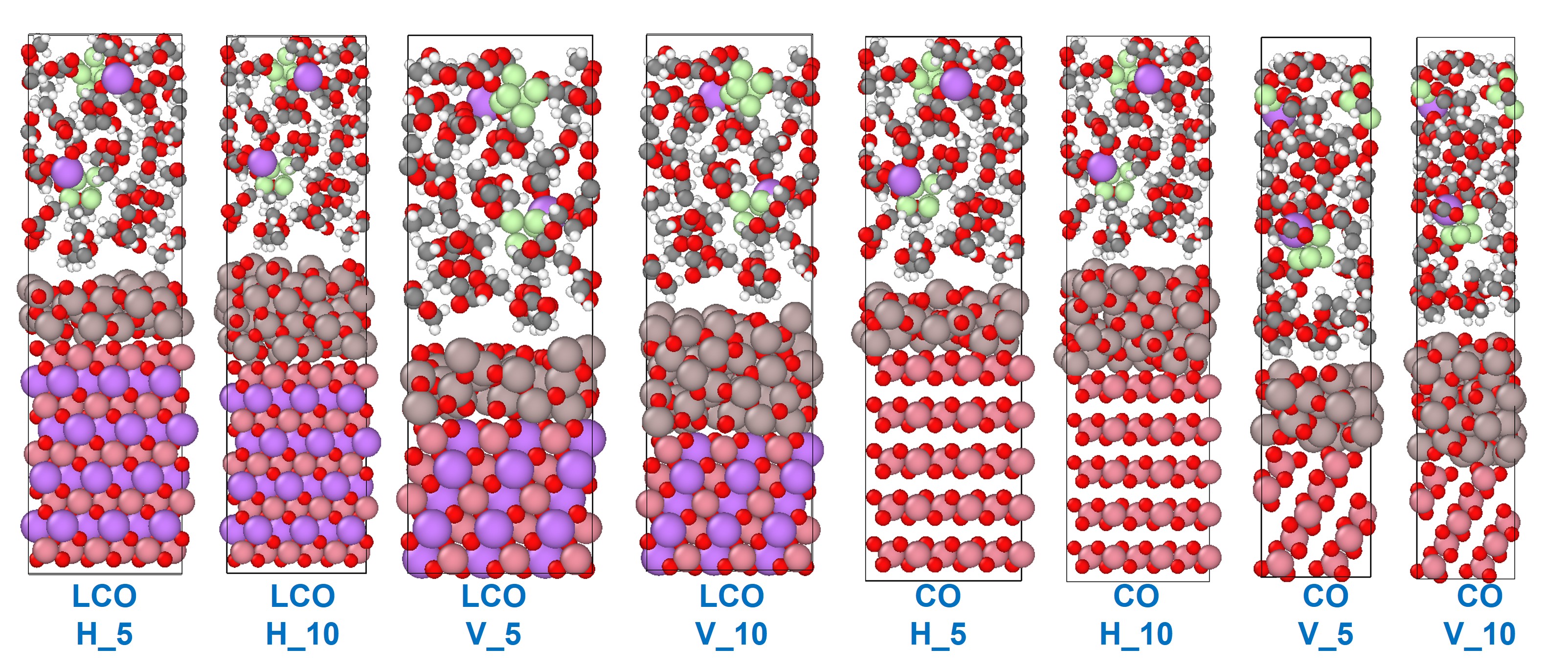}
    \caption{The four distinct systems with lithiated Co-rich cathodes and LiBF$_4$ as the salt in the electrolyte and Li channel orientations (H-horizontal, V-vertical) with two Al$_2$O$_3$ coating thicknesses (5~\AA~and~10~\AA). The top region consists of the electrolyte molecules (with LiBF$_4$ as the salt), the cathode at the bottom, and the Al$_2$O$_3$ coating at the cathode-electrolyte interface. Systems are not to scale. Co-pink, Li-purple, O-red, F-green, H-white, C-grey, Al-light brown}
    \label{fig:1_All_Systems}
\end{figure} 

\par Eight distinct systems (4 lithiated and 4 delithiated) were used to model the interactions between 1:1 EC:DMC solvent with  LiBF$_4$ salt on both a lithiated and delithiated LiCoO${}_{2}$ cathode, as shown in Figure~\ref{fig:1_All_Systems}. We identify these systems using the abbreviation ``V'' for systems with vertical channels and ``H'' for systems with planar (horizontal) channels ((012) and (001) planes, respectively). The thickness of the Al$_2$O$_3$ coating on the cathode slab will be indicated by the number that follows the channel orientation identifier. For example, system ``H\textunderscore5'' refers to the system with planar Li channels and a 5~\AA\ thick Al$_2$O$_3$ coating. Two separate trajectories were run for each system to vary the starting configurations, resulting in sixteen simulations. AIMD simulations were performed at 450~K with 1-fs time steps using the canonical (NVT) ensemble \cite{Lee2016} and were run upwards of 10~ps. This approach is typical for AIMD to accelerate the observation of reactions and system equilibration. Trajectories were tested for stability (nonphysical and rapid breakdown of components) up to 750~K during the design phase, and similar systems have been tested up to 800~K in recent work \cite{Joshua2019,Leung2012,zhang_stability_2020,young_preventing_2021,peiris_electrolyte_2024}. This work uses a temperature regulated at 450~K as a rational approach to balance the chemical stability of the system under consideration and achieve a sensible trajectory under the simulated timescales. We note that the electrolyte systems were chemically stable and free from spurious effects related to temperature regulation. All AIMD calculations were performed with a 400 eV plane wave energy cutoff using a $\Gamma$-centered \textit{k}-point. Van der Waals interactions were included using the Grimme D3 approach (PBE-D3) with Becke-Johnson damping \cite{Grimme2010}.

\subsubsection{Cathodes}
\par The layered cathodes were selected such that they represent a fully lithiated (discharged) LiCoO${}_{2}$ (LCO) or a fully de-lithiated (charged) CoO${}_{2}$ (CO) state. The planes (012) and (001) of the LCO and CO cathodes were used at the electrolyte-cathode interface, ensuring the periodicity of the cathode slab in two directions while serving as an active but stable interface for interactions with the electrolyte (Figure~\ref{fig:1_All_Systems}) \cite{Kang2015}.

\par Although our investigation is driven by the high-energy-density attributes of Ni-rich NMC materials, the limited cathode size makes it unfeasible to model the trace amounts of Mn and Co directly. Therefore, we have chosen LiCoO$_2$ as a representative end member within the family of NMC-type oxides. In addition, the fully delithiated phase CoO$_2$ is incorporated to distinguish the electrochemical reactions at different states of battery charge.
\par By taking advantage of the periodic boundary conditions of our VASP simulation cells, we model two electrolyte-cathode interfaces within a single simulation cell. This is achieved by constructing alternating layers of cathode and electrolyte materials at open-circuit voltage (OCV), a method consistent with previous studies that use similar models without vacuum regions \cite{Joshua2019,zhang_stability_2020,young_preventing_2021,le_determining_2017}. Our primary focus is to investigate the interfaces between the cathode and the electrolyte, particularly since the potential presence of a vacuum region can influence these interactions.
\par Analysis of the planar averaged local potential reveals that the potential remains periodic within the cathode, transitions in a smooth and continuous manner into the electrolyte and retains a consistent profile throughout the electrolyte region (see Figure~\textcolor{blue}{S2}).

\subsubsection{Electrolytes}

\par In the Supporting Information (SI), Table~\textcolor{blue}{S1} summarizes the species and atom counts employed in constructing the electrolyte systems. For each AIMD simulation, fourteen pairs of EC and DMC molecules were randomly allocated. Additionally, two pairs of LiBF${}_{4}$ salt were incorporated: one pair was positioned adjacent to the cathode surface, while the other was placed in the central region of the electrolyte to serve as a reference point for ion behavior. The configuration of these CIP salt pairs is illustrated in Figure~\ref{fig:1_All_Systems}.
\par The system volume was determined by equating the experimental 1 mol$\cdot$m$^{-3}$ EC/DMC concentration with the molecular concentration in the simulation. The requisite number of molecules was randomly distributed within the cell, after which the system underwent DFT relaxation prior to its integration with the cathode. To account for Coulombic repulsion, Van der Waals radii for oxygen were added on both sides of the cell. This approach yielded an electrolyte region with a density of approximately 1.28 g$\cdot$cm$^{-3}$ matching the experimental value (Sigma Aldrich \#746711).

\subsubsection{Coating and cell assembly}
\begin{figure*}[htbp]
	\centering
	\includegraphics[scale=0.55]{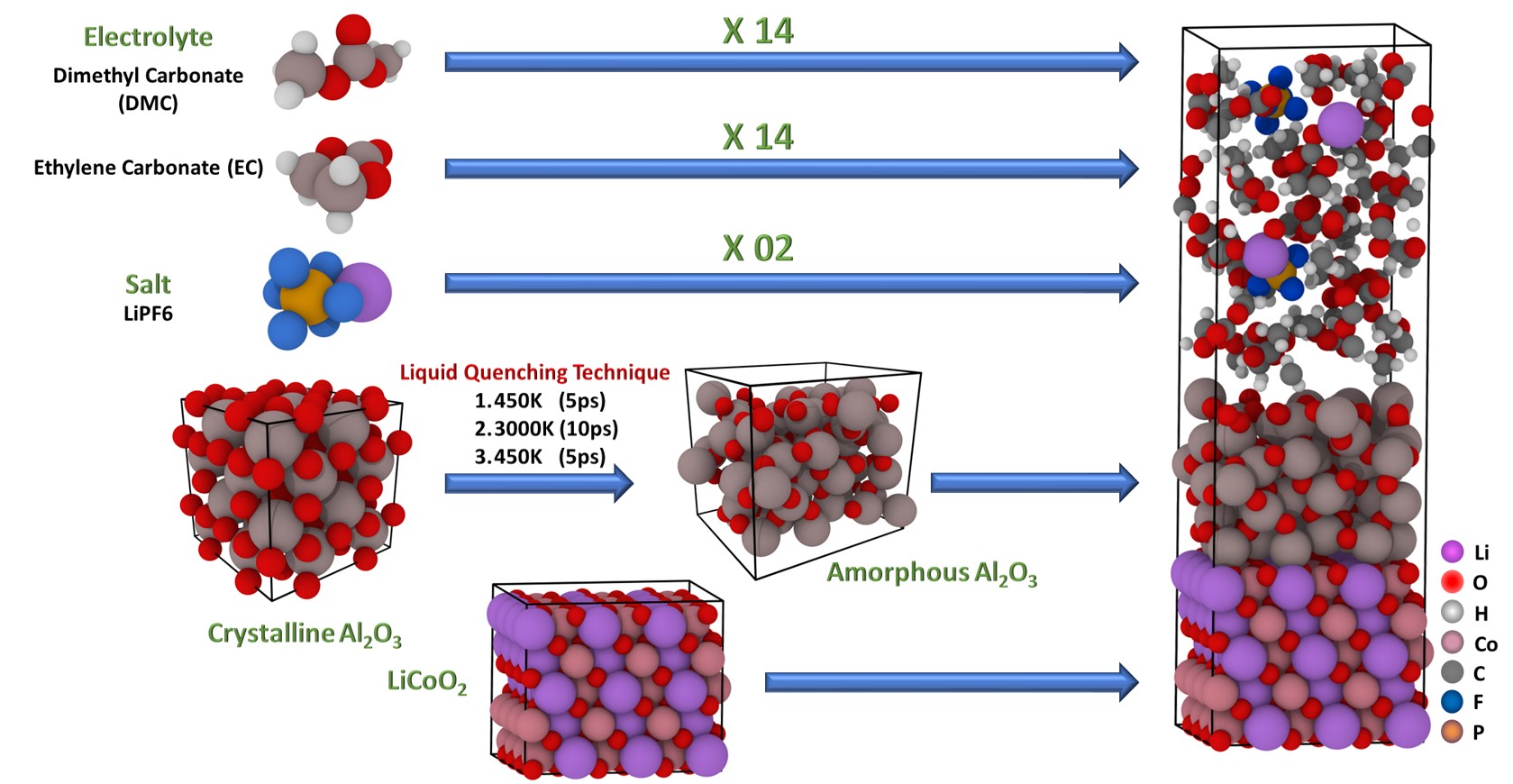}
	\caption{A cell consists of three regions: electrolyte, cathode, and coating. The individual regions were designed and relaxed separately and assembled into one system.}
	\label{fig:2_Assembly}
\end{figure*}

Using the quenching method, a bulk crystal cell of Al$_2$O$_3$ was prepared into an amorphous slab of appropriate dimensions for each cathode system. To address the loss of periodicity in the cell on the top and bottom surfaces, the amorphous slab was pre-relaxed before being placed on the top surface of the prepared cathode slab. The coating settled on the cathode surface during the equilibration of the system, which consisted of the electrolyte, coating, and cathode (see Figure~{\ref{fig:2_Assembly}}). 

\par The equilibration of the systems is critical in MD simulations and was verified by monitoring the energy and temperature regulation of a given system with respect to time. Additionally, the electronic convergence of the calculation at each ionic step was closely monitored.

\par The analysis of the trajectories, including but not limited to the observations of reactions, disintegration process of molecules, and inter-atomic distances, was done using the cluster analysis and color-coding options in the Ovito program. Bond distances were calculated using a Python script developed in-house. The results were initially cross-checked with the Ovito software to verify accuracy.

\subsection{Cleavage energy}
\begin{figure}[!h]
\centering
{\includegraphics[scale=0.35]{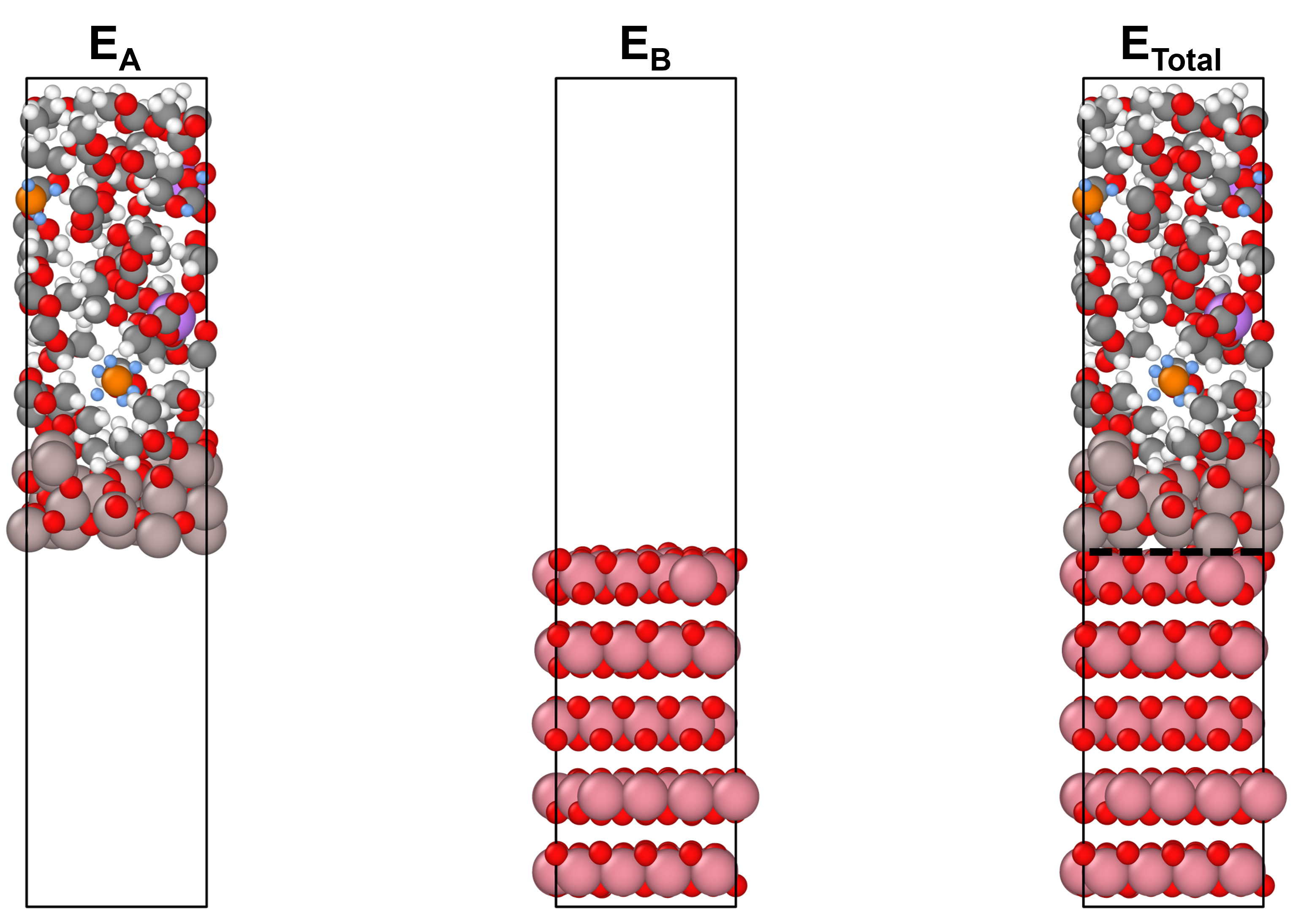}}
\caption{Graphical representation of how the energy values in Eq.~\ref{cleavage} are evaluated, with (a) $E$\textsubscript{A}, (b) $E$\textsubscript{B} and (c) $E$\textsubscript{Total}. An example of a cleavage location is shown as a dashed line (black).}
\label{Cleavage_Energy}
\end{figure}

For calculations of the cleavage energy (\textit{i.e.}, the energy required to separate the two parts of the interface), we use the Dupré equation \cite{Jung2013,Woodcox2022},
\begin{equation}
W\textsubscript{sep}=\frac{E\textsubscript{A}+E\textsubscript{B}-E\textsubscript{Total}}{S} ,
\label{cleavage}
\end{equation}
where $E$\textsubscript{A} and $E$\textsubscript{B} are the energies of the A and B regions which compose the interface, $E$\textsubscript{Total} is the total energy of the system, and $S$ is the area of the interface. The values of $E$\textsubscript{A} and $E$\textsubscript{B} are obtained by removing the complementary region from the simulation cell and replacing it with vacuum, as shown in Figure~\ref{Cleavage_Energy} \cite{Jung2013,Woodcox2022}. This value quantitatively assesses the material's overall strength at the cleavage point. An additional vacuum layer of 30~\AA~is added to each of these snapshots in the \textit{z}-direction before the energy calculations to prevent interactions between periodic images.

Mechanically controlled break junctions (MCBJs) are often used to explore molecular electronics \cite{DaSilva2004,Lonjon2010,Tavazza2010,Cortes-Huerto2013,Li2019,Ivie2021}. In our previous work, we used AIMD to simulate MCBJs at solder interfaces for comparison to changes in $W$\textsubscript{sep}, and we employ that method in this work to assess interfacial strength \cite{Woodcox2022}. This technique provides a more realistic depiction of interfacial dynamics under stress, but it comes at a notably higher computational cost \cite{Woodcox2022}.
The top 50\% of atoms in the Al$_2$O$_3$ coated surface and all atoms in the bottom layer of the cathode (exposed to the electrolyte) were frozen during this process to maintain stability in the coating and provide an anchor point for the stretching, respectively. To simulate the MCBJ process, the atomic positions of the Al$_2$O$_3$ layer along the $z$-axis were manually shifted by 0.1~\AA~using a 100~fs time interval. This would allow sufficient time for all atoms to respond based on the new forces/stresses introduced by moving the topmost layers. All calculations were run using the canonical ensemble (NVT) at 750~K to ensure we would not approach the melting point of the simulated material. Employing both the $W$\textsubscript{sep} and MCBJ methods allows for rapid exploration of various systems, ensuring reliable results and comprehensive analysis of crucial material components.

\section{Results and discussion}
\subsection{Surface Chemistry} 

EC was the most chemically active electrolyte species on the interfaces, starting with the exposed cathode-electrolyte interface. The exposed cathode surfaces showed significant activity in breaking down the EC molecules compared to the coated cathode surface. The most common form of electrolyte degradation was the initial rapid dehydrogenation of the EC molecules, which was particularly frequent on the delithiated surfaces. Of the two surfaces we simulated, horizontal (001) and vertical (012) channeled systems, vertical channeled surfaces were more reactive. This is expected since the horizontal channels have no dangling bonds in the layer exposed to the electrolyte, while the (012) surface inherently contains atoms that can be under-coordinated, opening the surface for increased chemical activity (\textcolor{blue}{Figure~S1)} \cite{sharifi-asl_facet-dependent_2017,kang_first-principles_2015,genreith-schriever_oxygen_2023}. The (012) and the (014) facets have previously been shown to be vulnerable to thermal degradation as well \cite{sharifi-asl_facet-dependent_2017}. 
    
\par Once dehydrogenated, the EC decomposition reaction was observed to proceed in several different ways (\textcolor{blue}{Figure~S1~A-B}). It could either continue to remove the other hydrogens from the system and stay bound to the cathode surface via its ether oxygen or undergo further reactions on the surface \textcolor{blue}{(Figure~S1 B)}. The transferred hydrogens form hydroxyl on the cathode surface, with the propensity to form H$_2$O by combination with another hydroxyl by H `hopping' on the surface oxygens. Upon the complete breakdown of an EC molecule, the frequent end products have been observed to be CO$_2$ molecules, CO$_3$$^{2-}$, and HCO$_3$$^{-}$ species \cite{peiris_electrolyte_2024}.  

\par Compared to EC, DMC did not engage in rapid or extensive reactions. Dehydrogenation at the delithiated interfaces was the most frequent reaction form seen with DMC molecules \textcolor{blue}{(Figure~S1-C)}. These DMCs did not experience a further breakdown within the rest of the trajectory. The methyl oxygen that loses the H would typically stay bound to the cathode surface. On one occasion, given the surface instability of the oxygen species present on the exposed (012) cathode surface, it resulted in the formation of methyl glyoxal species. Once a few of these DMC and EC molecules had broken down on the surface, we observed what is understood to be the early signs of a CEI layer formation. Deactivated or surface-bound electrolyte species significantly slowed surface interactions with other electrolytes and introduced a certain degree of passivity to the exposed cathode surface. 

\par There was a considerable formation of O$_2$ molecules from the delithated, (012) cathode systems (\textcolor{blue}{Figure~S3}). This observation was consistent with the studies we have performed with LiNiO$_2$ cathodes, where we reported on the formation of singlet oxygen from radical oxygen species on a similar (012) cathode surface \cite{peiris_electrolyte_2024}. The O$_2$ molecules, once formed, can move away from the surface, resulting in a non-reversible form of surface degradation. Interactions with the electrolyte during that formation can lead to electrolyte degradation, as shown in \textcolor{blue}{Figure~S1}.
    
\par The reactivity of the coated surface was observed to be minimal in the trajectories. In a single instance, the EC molecule underwent deprotonation at the Al$_2$O$_3$ coated surface, where an Al$_2$O$_3$ oxygen is involved in the process \textcolor{blue}{(Figure~S5)}. The proton immediately transfers to a nearby oxygen atom, allowing the methyl carbon to bond with the surface oxygen. After approximately 2-3~ps, a `bridging' is observed, where a ring opening of the surface-bound EC leads to a carbonyl oxygen bonding to a surface Al atom. Although complete degradation of the EC is not achieved within the time frame of this trajectory, we anticipate that further bond cleavage will result in the formation of either a carbonate species or CO$_2$, depending on the location of the subsequent bond scission.

\par O$_2$ formation on the coated surface exposed to the electrolyte was observed but was minimal relative to the exposed delithiated (012) surfaces. The few instances of O$_2$ emission came from the combination of nearby under-coordinated oxygen atoms and were observed only in the initial steps of the trajectory ($\sim$1-2~ps). Similar to the O$_2$ emissions from the cathode exposed to the electrolyte, the O$_2$ would move away from the surface once formed. 
    
\subsubsection{Solvation dynamics of the salt}

\par
\begin{figure}[htbp]
    \centering
    \includegraphics[scale=0.55]{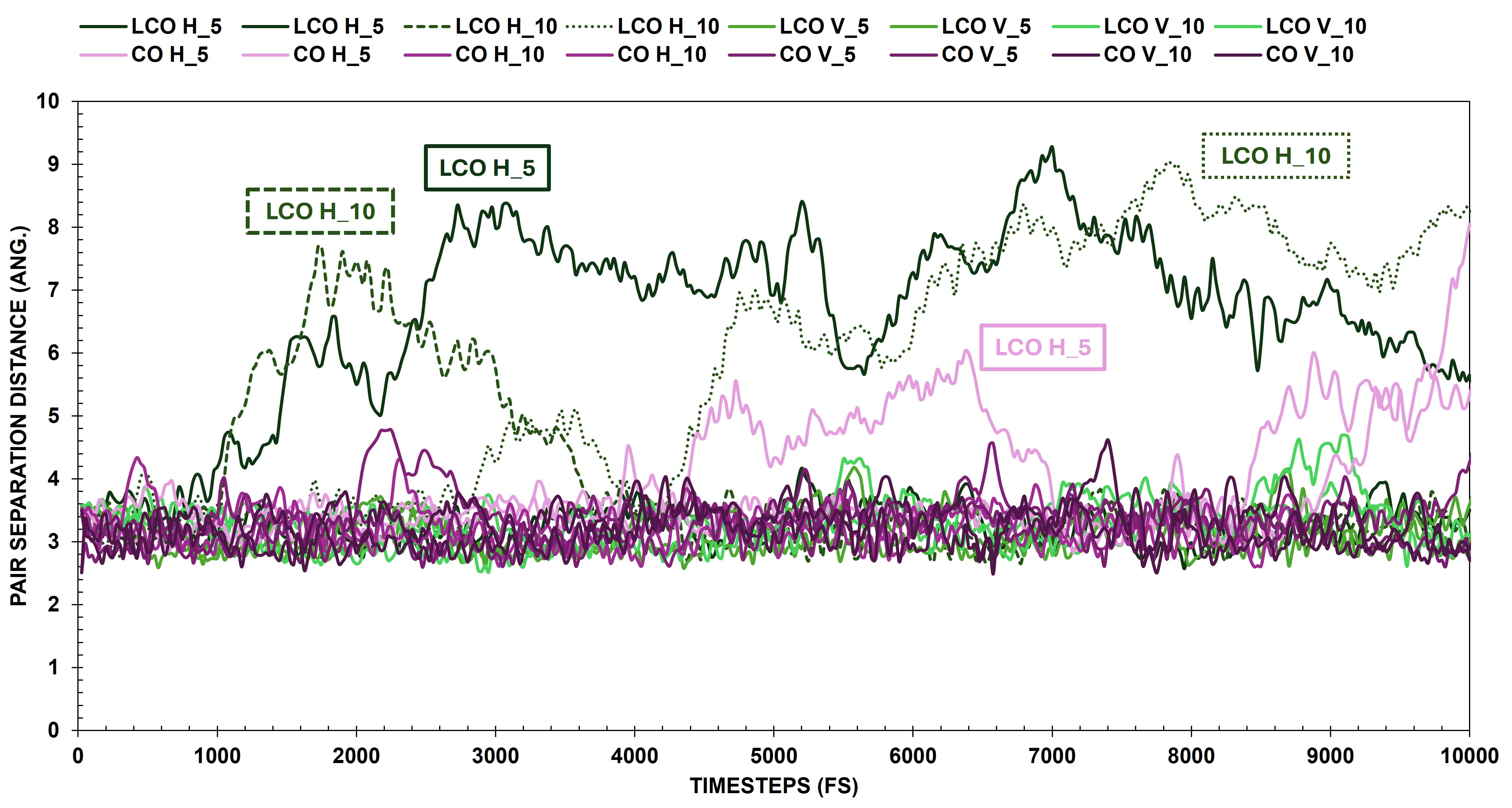}
    \caption{The cation-anion separation distance (as measured from Li to P) as a function of time for all AIMD trajectories. Legend: V for (012) plane, H for (001) plane.}
    \label{SaltDistances}
\end{figure}

 The initial structure of the salt was contact ion pair (CIP) to reflect the strong electrostatic interactions between ions in close proximity, which is common in high ionic strength solutions. This approach allowed for efficient exploration of ion pairing dynamics, solvation effects, and ion dissociation processes, providing a basis for understanding the solvation dynamics. We monitored the persistence of the CIP structure in all the trajectories, and the Li-P distance is shown in Figure~\ref{SaltDistances}. With the exception of one instance of anion-cation dissociation in the delithiated cathode, all observed separations occurred nearer to the lithiated cathode surface—a trend also noted in LiNiO$_2$ cathode-electrolyte interfaces \cite{peiris_computational_2023}. This behavior is attributed to a surplus of lithium at the surface, which reduces the tendency for the Li$^+$ cation to remain within its original solvation shell. In cases where the salts maintained a contact ion pair (CIP) configuration, the average separation between the Li cation and BF$_4^-$ anion consistently measured approximately 3.28~\AA~\textcolor{blue}{(Fig~S6)}. 

The environment around the cation of the salt was analyzed to elucidate the persistence of the initial solvation shell that forms around it. To identify the number of EC and DMC molecules participating in the shell, a cluster radius of 2.9~\AA~was used, and the number of species and their type were tracked throughout the trajectory. An example of such a plot is shown in \textcolor{blue}{Figure~S7} in SI. This was aided by filtering the trajectory for such clusters and used for visual verification to maintain accuracy \textcolor{blue}{(Fig~S8)}. We find that the LiBF$_4$ salt is typically surrounded by three electrolyte molecules when it maintains the CIP structure. This is in agreement with previously reported experimental and computational values for the solvation of LiBF$_4$ \cite{peiris_computational_2023}. It is understood that a Li cation typically maintains a tetrahedral solvation structure in organic electrolytes, allowing up to 4 counterparts. As long as the CIP structure prevails, one of these positions will be occupied by the anion, thus reducing the available spots for solvation to 3, similar to what we observe in our trajectories \textcolor{blue}{(Table~S2 in SI)}. Looking into the composition of the solvated Li cation, we find that EC molecules are more dominant in participating in the cell compared to the linear DMC molecules by $\approx$78~\%. However, there seems to be no net transfer in charge from the Li cation or the anion during the breakdown of the CIP structure \textcolor{blue}{(Figure~S9)}. 

\subsubsection{Effect of the coating on the charge of TM atoms}
\begin{figure}[t]
    \centering
    \includegraphics[scale=0.5]{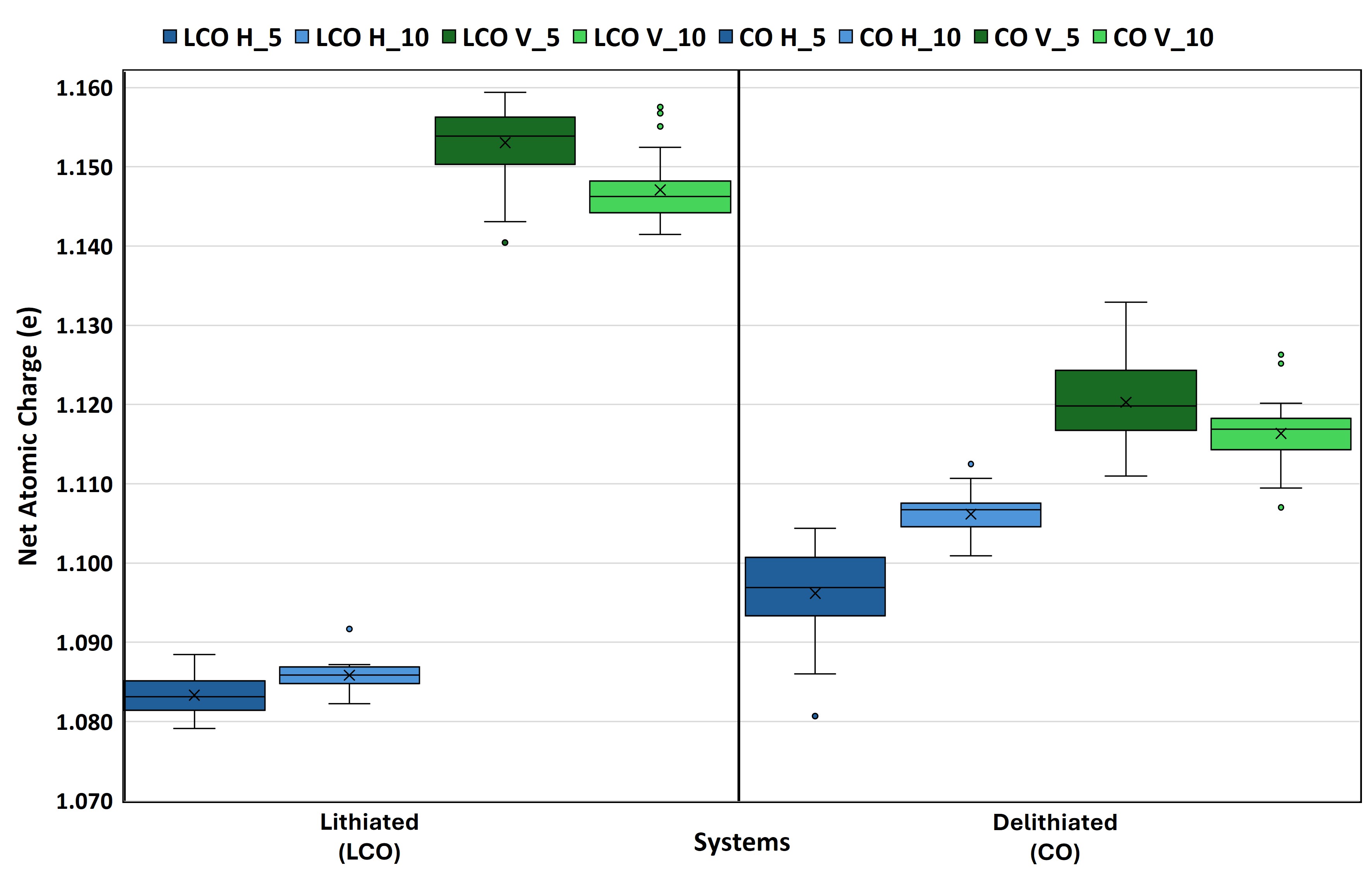}
    \caption{The variation of the averaged net atomic charge on Co atoms for all the systems considered in this study. Each colored pair shows a grouping of the systems by lower (5~\AA) and higher (10~\AA) coating thicknesses and by the orientation of the Li channel (horizontal (H) and vertical (V)). The first two pairs from the left show systems with lithiated cathodes (LCO), and the next two show delithiated systems (CO). Legend: V for (012) plane, H for (001) plane.}
    \label{BW-All}
\end{figure} 

We extracted the net atomic charges for each Co atom to probe how facet orientation, lithiation state, and Al$_2$O$_3$ coating thickness influence the electronic environment of Co using the DDEC6 method on AIMD snapshots extracted every 200~fs. We then analyzed the variation of these Co net atomic charge values for each layer and the total system (Figure~\textcolor{blue}{\ref{BW-All}} and Figure~\textcolor{blue}{S10}). In this convention, a more positive net atomic charge means Co has lost electron density to its surroundings, while a less positive value indicates net electron density accumulation on Co.

Figure~\ref{BW-All} compares the distributions of Co net atomic charges across eight systems: lithiated and delithiated forms of the (001) and (012) facets, each coated with either 5~\AA\ or 10~\AA\ of Al$_2$O$_3$. In the lithiated state, the (001) facet shows the lowest median Co charge, indicating minimal electron withdrawal, while the (012) facet shows a higher median value, consistent with significant electron density loss from Co. Delithiation shifts both facets to even higher positive charges, reflecting further electron loss. Increasing the Al$_2$O$_3$ coating from 5~\AA\ to 10~\AA\ slightly increases the Co charge (by $\sim$0.002~$e$) on the (001) facet, but decreases it (by $\sim$0.006~$e$) on the (012) facet. These trends suggest that both facet orientation and coating thickness modulate interfacial electron redistribution. Although absolute DDEC6 net atomic charges carry an uncertainty of about ±0.01~e, the trends are consistent throughout the trajectory and align with our prior results for LiNiO$_2$ \cite{peiris_electrolyte_2024}, supporting the qualitative conclusions drawn from these data.


Layer-resolved analysis reveals a distinct interfacial effect on Co charge states depending on whether the atoms were exposed to the electrolyte or the Al$_2$O$_3$ coating (Figure~S10). In the lithiated (001) and (012) electrodes, Co atoms at the electrolyte interface exhibited a greater increase in positive net charge, indicating more substantial electron depletion, compared to those beneath the Al$_2$O$_3$ layer. When fully delithiated, this trend reversed: the coated Co atoms experienced more pronounced electron loss than those facing the electrolyte, indicating a lithiation-dependence of charge transfer at the interface.

We can summarize the observations from Figure~\textcolor{blue}{\ref{BW-All}} as, (i) the lithiated (001) surface exhibits minimal Co charge transfer while the lithiated (012) surface shows maximal charge redistribution; (ii) upon delithiation these facet differences become less extreme; and (iii) a thicker Al$_2$O$_3$ layer increases charge transfer on (001) but suppresses it on (012).

The (001) basal plane of LiCoO$_{2}$ is relatively stable and “chemically saturated,” lacking dangling bonds, unlike the (012) facet (sometimes indexed as (104) in rhombohedral coordinates) that exposes under-coordinated atoms. \textit{Ab initio} and spectroscopic studies have shown that Co on different facets has different spin/valence states \cite{hong_electronic_2019}. In particular, Co$^{3+}$ on the polar (001) surface remains in a low-spin state (t$_{2}$g$^{6}$, like bulk Co$^{3+}$) and has an electronic structure similar to bulk LiCoO$_{2}$. In contrast, Co on the more open (012) surface can adopt mixed or high-spin states due to reduced ligand field (fewer O neighbors) \cite{hong_electronic_2019}. This means the (012) facet has more reactive Co–O bonds and a higher tendency for electronic polarization. The low-spin Co$^{3+}$ on (001) is already stable and less prone to additional charge transfer, whereas mixed-spin Co on (012) has unpaired 3d electrons that can be more readily redistributed or shared in new bonds.

When the Al$_2$O$_3$ coating is applied, it bonds differently to these facets. ALD-Al$_2$O$_3$ is amorphous but will bind via Al-O bonds to available surface oxygens or lithium sites. On the (001) surface, which is densely packed, there are fewer dangling bonds for the coating to attach to. Even if the (001) surface is Li-terminated, the Li$^{+}$ can donate some electron density to surface O$^{2-}$, resulting in relatively little unsatisfied charge for the Al$_2$O$_3$ to attract. Therefore, the coating may essentially form on top of a stable CoO$_{2}$ layer with minimal direct Co–O–Al bonding. As a result, the charge on Co is only weakly perturbed on (001) in the lithiated state (\textcolor{blue}{Figure~S11}). In contrast, the (012) surface presents under-coordinated O (and possibly exposed Co). The Al$_2$O$_3$ will more strongly chemisorb – for example, by forming interfacial Al-O-Co linkages. Previously, DFT calculations (on analogous Ni-rich cathodes) have indicated that Al from the coating can coordinate to surface oxygen that is also bonded to transition-metal cations \cite{chen_identification_2025}


In the delithiated state, the difference between the (001) and (012) surfaces’ behavior with Al$_2$O$_3$ is less pronounced - the trends become “intermediate,” meaning the (001) no longer has a trivial change, and (012) no longer has an extreme change, but both show a moderate charge redistribution. Mechanistically, this is because Co on both facets is now in a higher valence state, eager to gain electron density. In essence, the surface Co charge distribution in the delithiated (001) surface becomes more similar to (012) in that it experiences a significant interfacial charge shift (where previously it did not). Conversely, the (012) facet in the lithiated state had already undergone a considerable charge transfer with the coating; upon delithiation, Co oxidation cannot increase much further and leads to structural changes (surface reconstruction, O$_2$ formations (Figure\textcolor{blue}{~S3}), and diffusion of Al into empty channels in our simulations (Figure\textcolor{blue}{~S12})). Therefore, delithiation puts both (001) and (012) in a state where the Al$_2$O$_3$ layer significantly interacts with Co, leading to intermediate levels of charge transfer for both facets (as opposed to the lithiated case, where (001) was almost unaffected and (012) was highly affected). Looking at the charge transfer from Co atoms at each layer on (001), this becomes more evident (Figure~\textcolor{blue}{S11}). The layers that are exposed to the electrolyte see more charge transfer from TM when lithiated in each facet, while when fully delithiated, the layer dependence of the transferred charge becomes similar for both facets.

A thin ($\sim$5~Å) Al$_2$O$_3$ film primarily affects the most reactive sites (strong effect on (012), weak on (001)), whereas a thicker ($\sim$10~Å) film enforces a more uniform, strained, and polarized interface (increasing the effect on (001) and saturating/covering the effect on (012), see Figs.~\textcolor{blue}{S10-S11,S13}). Thicker coatings also act as better barriers, which can reduce localized charge-transfer intensity on reactive facets. Previous DFT-based interface studies show that ultrathin films can have a different chemistry from slightly thicker ones. A monolayer of ZnO on LiCoO$_{2}$ was found to react and get etched by HF, whereas a few-nm-thick ZnO layer resisted complete dissolution \cite{tebbe_mechanisms_2015}.

\subsubsection{Mechanical properties of the coating}
Having examined the reactivity and solvation characteristics of the coated and uncoated cathodes in the two end states of lithiation, we now examine the mechanistic effects of the oxide coating on the cathode surface.
We start by calculating the energy required to cleave the cathode-coating system at points of interest. We then examine the trends between delithiation patterns and how they influence the mechanical properties of the coating during the charging and discharging of a cathode. 


\begin{figure*}[ht]
\includegraphics[scale=0.6]{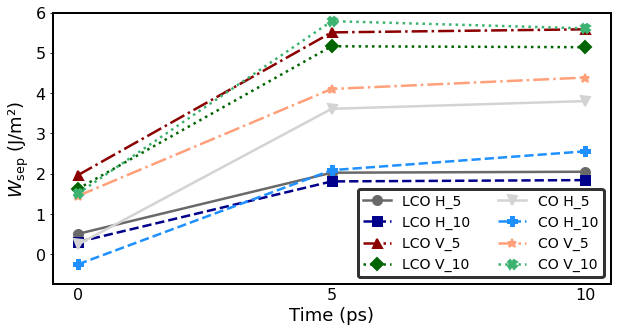}
\caption{Cleavage energy ($W$\textsubscript{sep}) at the cathode-Al$_2$O$_3$ interface at 0, 5, and 10~ps for each of the eight structures studied in this work.}
\label{CE_time}
\end{figure*}

An initial assessment of $W$\textsubscript{sep} at the cathode-Al$_2$O$_3$ interface was performed to confirm that significant changes in the $W$\textsubscript{sep} were no longer occurring in $W$ \textsubscript{sep}.
This would serve as validation that sufficient equilibration of the mechanical processes had taken place. The results of this test are shown in Figure~\ref{CE_time}, where $W$\textsubscript{sep} for each system shown in Figure~\ref{fig:1_All_Systems} were calculated in the initial step ($t$~=~0~ps), an intermediate step ($t$~=~5~ps), and the final step ($t$~=~10~ps) of the AIMD trajectories.
There are significant increases in $W$\textsubscript{sep} for all systems from $t$~=~0~ps to $t$~=~5~ps.
While the overall trends are the same in the two datasets, the transition from the DFT relaxed structure to the AIMD equilibrated interface predicts vastly different values for $W$\textsubscript{sep}. Minimal changes to $W$\textsubscript{sep} occur from $t$~=~5~ps to $t$~=~10~ps, indicating sufficient equilibration of the interface. For both the lithiated and delithiated structures, we find that the vertical channels provide an appreciable benefit to the interfacial strength in terms of $W$\textsubscript{sep}. In general, we find that increasing the thickness of the Al$_2$O$_3$ coating will decrease $W$\textsubscript{sep} at the cathode. The additional Al$_2$O$_3$ with the delithiated vertical channels (V\textunderscore10) shows a substantial benefit in this analysis. Due to the amorphous nature of the coating, the cathode-Al$_2$O$_3$ interface was the initial location to assess for weakness within the system. However, this is not the only potential source of weakness.

Above, we have focused on the cathode-Al$_2$O$_3$, but to further assess the effects of the coating, we calculated $W$\textsubscript{sep} within each of the empty cathode layers of the CO H\_5 structure. For this structure, we find that the interface is appreciably stronger ($W$\textsubscript{sep}~=~3.8~J/m$^2$) than the interlayer cathode spaces, which have $W$\textsubscript{sep}~$<$~0. These negative values indicate that separation within the structure of the cathode itself requires minimal amounts of energy. In certain layered materials, it has been determined that negative values for $W$\textsubscript{sep} are due to weak van der Waals interactions \cite{prodanov2010computational}. Alternatively, induced strain within materials can lead to energy release upon cleavage because it allows for the atoms to relax into a lower-energy state \cite{gould2009van,mcguire2017magnetic}. In our calculations of $W$\textsubscript{sep}, the atoms are not allowed to relax (having only performed self-consistent calculations for $W$\textsubscript{sep} to hold them in the same positions as the AIMD trajectories), indicating that weak van der Waals interactions are responsible for the negative values.

\begin{figure*}[ht]
\subfloat[LCO H\textunderscore5]{\label{Cutting_Lithiated}\includegraphics[width=0.5\columnwidth]{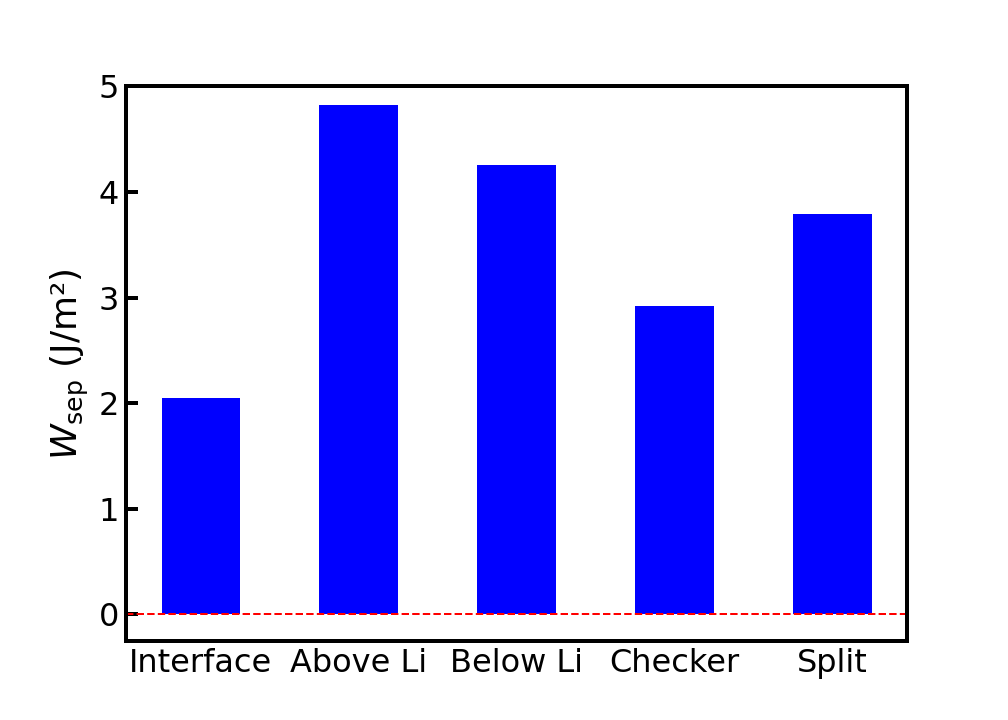}}
\subfloat[$W$\textsubscript{sep} vs. Li Concentration in top layer]
{\label{CE_Li}\includegraphics[width=0.5\columnwidth]{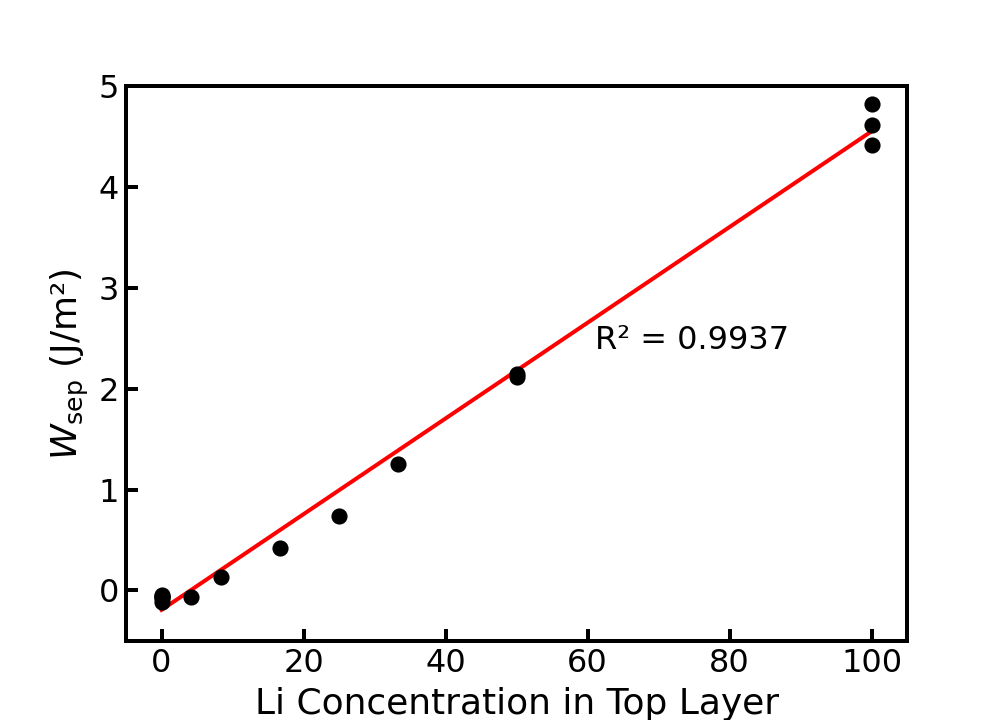}}
\caption{(a) $W$\textsubscript{sep} of the lithiated H$_5$ structure. We show $W$\textsubscript{sep} at the cathode-Al$_2$O$_3$ interface, and four different cleavage options within the top Li layer of the cathode. (b) $W$\textsubscript{sep} as a function of Li concentration in the top layer of the cathode, with a linear fit (red), as well as the R$^2$ value to show goodness of fit.
\label{CE_Cutting}}
\end{figure*}

There are additional choices for evaluation for the lithiated H\_5 structure because the possible cleavage locations around the Li atoms could yield notably different results. We have focused on the topmost layer of Li for our calculations, but chose four possible cleavage orientations within that layer: 1) Above all Li atoms, 2) below all Li atoms, 3) Checkerboard (selecting every other Li atom and allowing half of them to be separated with the top of the cathode and the other half to be separated with the bottom part of the cathode) and 4) Split (selecting Li atoms based on their position relative to the center of the cathode and allowing the left half of the Li atoms to be separated with the top of the cathode and right half to be separated with the bottom part of the cathode). 

The results of these cleavage options are shown in Figure~\ref{Cutting_Lithiated}. This variety of cleavage energies allowed us to test the energetic benefits of different types of cleavage.
The cathode–Al$_2$O$_3$ interface is the strongest region in the delithiated structure; it becomes the weakest in the lithiated cathode, despite having a relatively high $W$\textsubscript{sep}.

\begin{figure}[!htbp]
    \centering
    \includegraphics[scale=0.45]{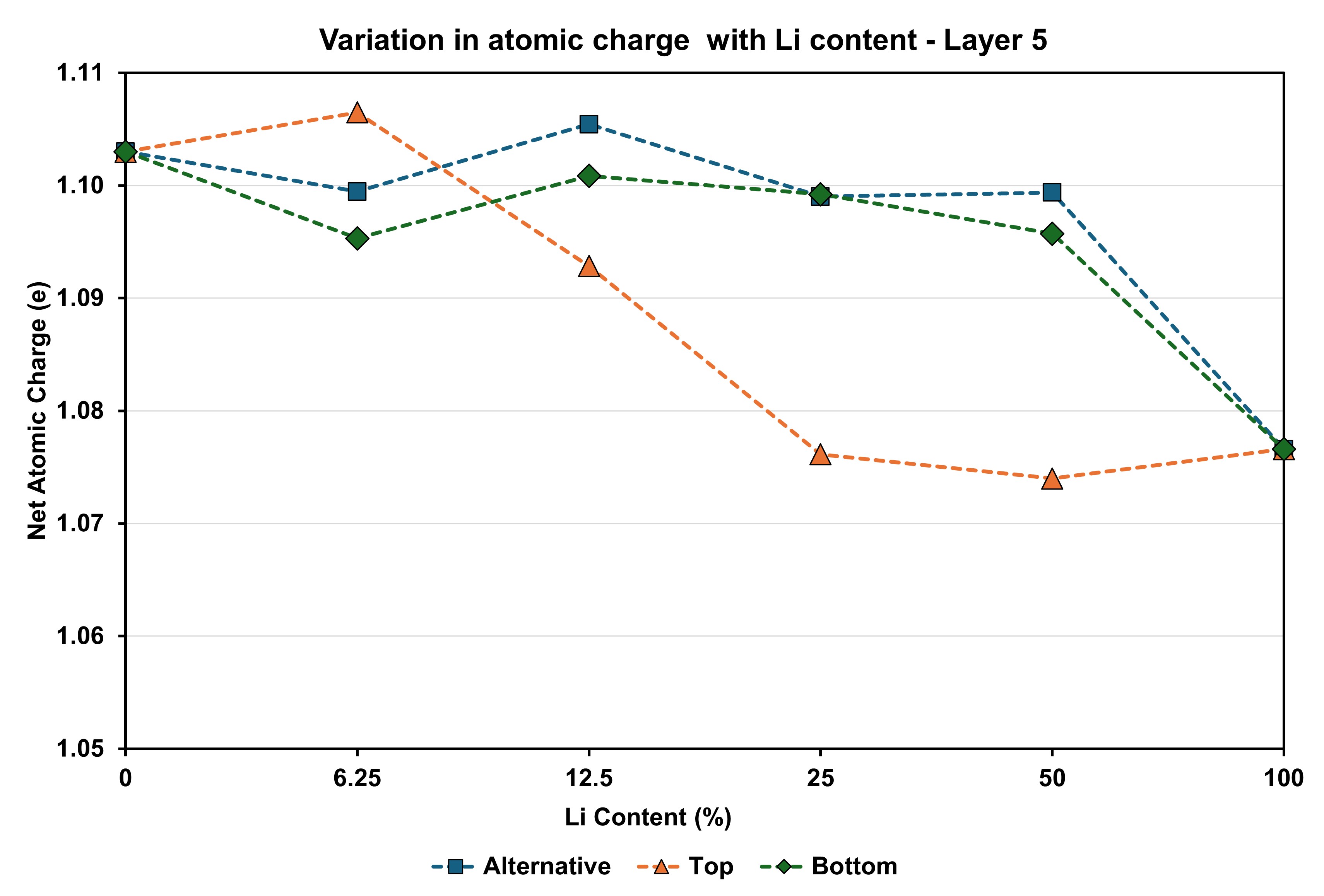}
    \caption{The change in the net atomic charge of the Co atoms located closest to the Al$_2$O$_3$ coating, grouped by the Li content in the cathode. The bars are colored according to the approach used for the removal of Li from the cathode.}
    \label{Layer5 charge}
\end{figure} 

To further demonstrate the relationship between $W$\textsubscript{sep} and Li concentration, additional AIMD trajectories were run with intermediate Li concentrations: 50~\%, 25~\%, 12.5~\%, and 6.25~\%. See \textcolor{blue}{Figs.~S14-S17} for the illustrations of each Li removal approach at each delithiation level. Multiple trajectories were run for each concentration because of the different relative concentrations of Li within each layer. For Li concentrations of 50~\% (\textcolor{blue}{Figure~S14}), 12.5~\% (\textcolor{blue}{Figure~S16}), and 6.25~\% (\textcolor{blue}{Figure~S17}), three relative orientations were used, while four different Li positions were used for a Li concentration of 25~\% (\textcolor{blue}{Figure~S15}). For each structure, their energy difference relative to the fully lithiated cathode is also included in \textcolor{blue}{Figs.~S14-S17}. Within these trajectories, each concentration contained an option for `Top', where the emphasis was placed on having as many Li atoms in the top layer as possible, and `Bottom', where the emphasis was to place as many atoms in the bottom layer as possible, and `Alternating', where the Li ions were distributed as evenly as possible throughout all layers of the cathode. While this is not an exhaustive list of all possible relative orientations of Li throughout the cell, it is representative of the variety of possible structures. The resulting $W$\textsubscript{sep} as a function of Li concentration in the top layer of the cathode is shown in Figure~\ref{CE_Li}. We find a direct correlation between the amount of Li in the top layer of the cathode and $W$\textsubscript{sep} with an R$^2$~=~0.9937 for the 15 AIMD trajectories tested in this analysis.

\begin{figure*}[!htbp]

\captionsetup[subfigure]{position=top}
\subfloat[MCBJ Energy]{\label{MCBJ_Energy}\includegraphics[width=0.75\columnwidth]{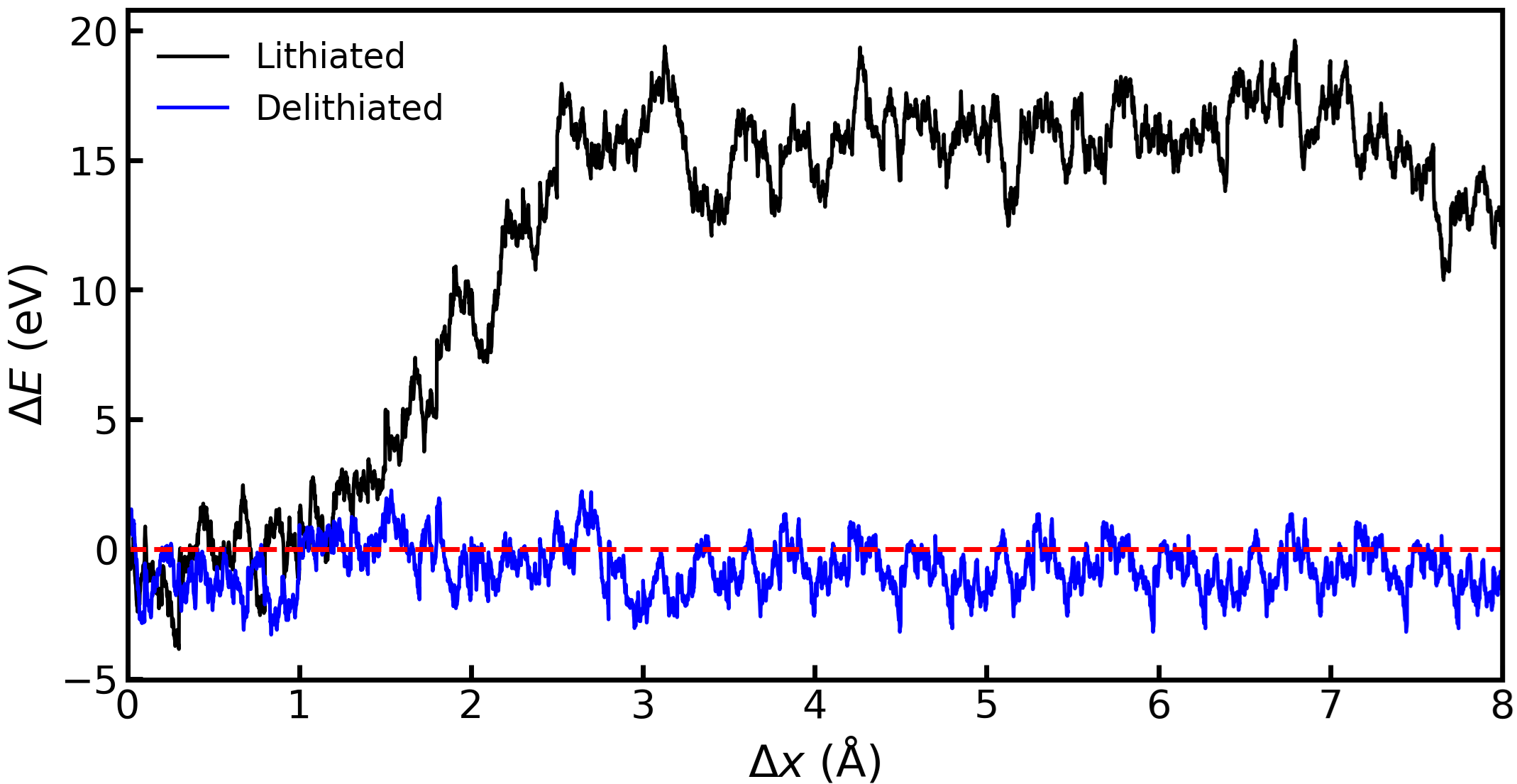}}\\
\vspace{0.5cm}
\captionsetup[subfigure]{position=bottom}
\subfloat[Lithiated $\Delta x$~=~2.0~\AA]{\label{MCBJ_B}\includegraphics[width=0.33\columnwidth]{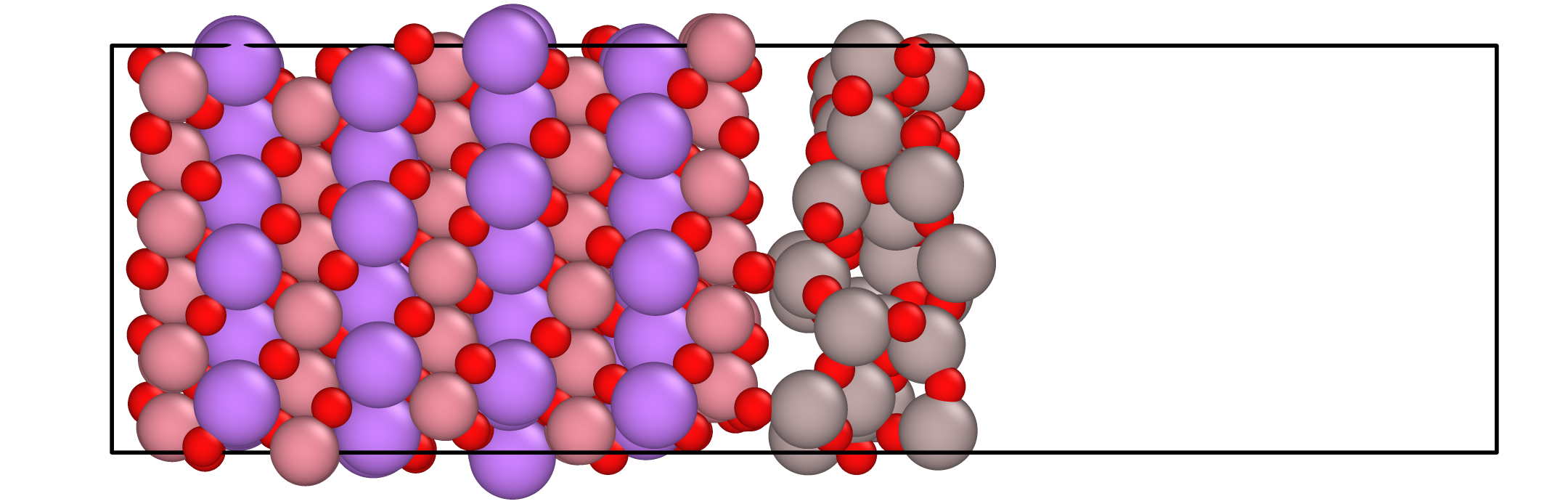}}
\subfloat[Lithiated $\Delta x$~=~4.0~\AA]{\label{MCBJ_C}\includegraphics[width=0.33\columnwidth]{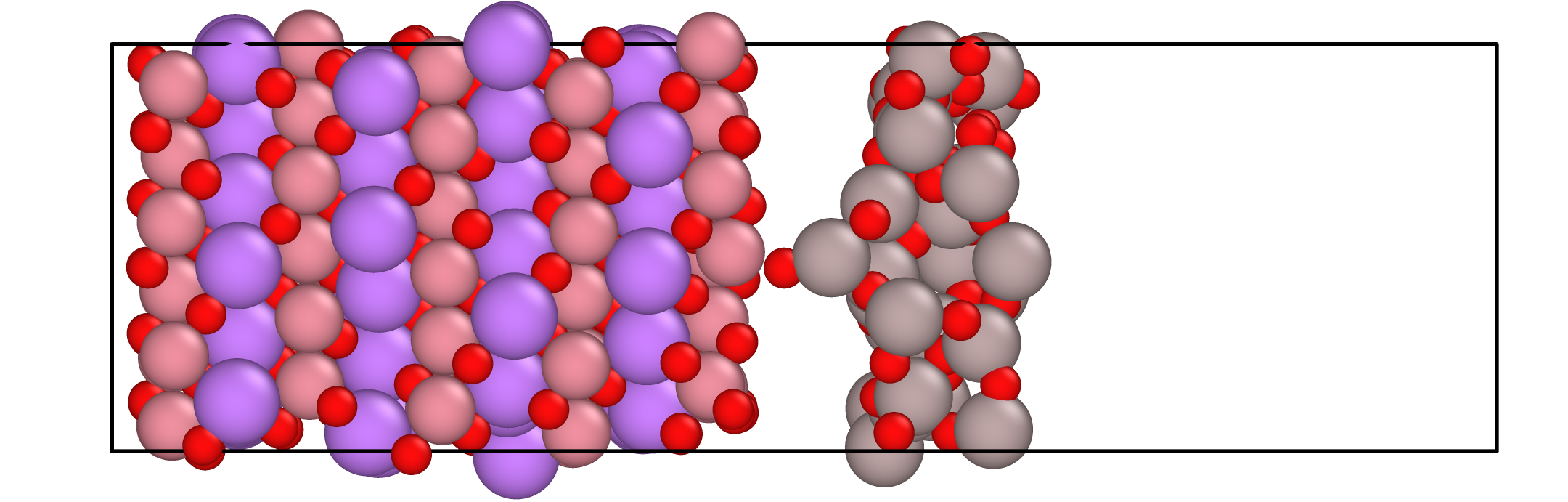}}
\subfloat[Lithiated $\Delta x$~=~8.0~\AA]{\label{MCBJ_D}\includegraphics[width=0.33\columnwidth]{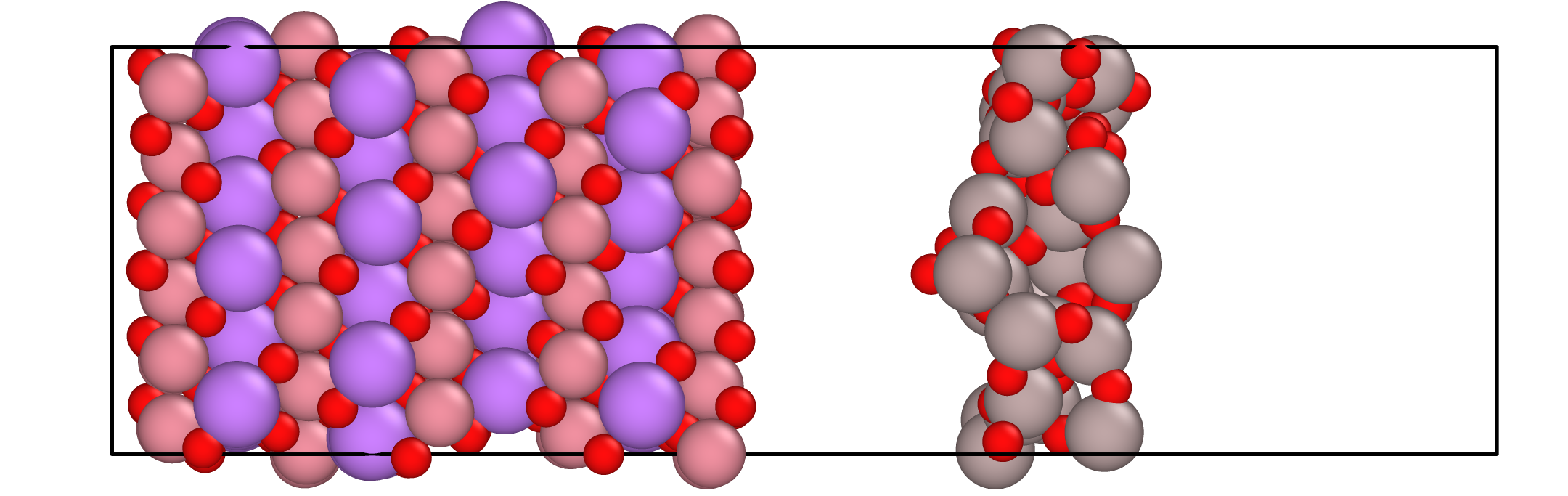}}\\ 
\vspace{10pt}
\subfloat[Delithiated $\Delta x$~=~2.0~\AA]{\label{MCBJ_E}\includegraphics[width=0.33\columnwidth]{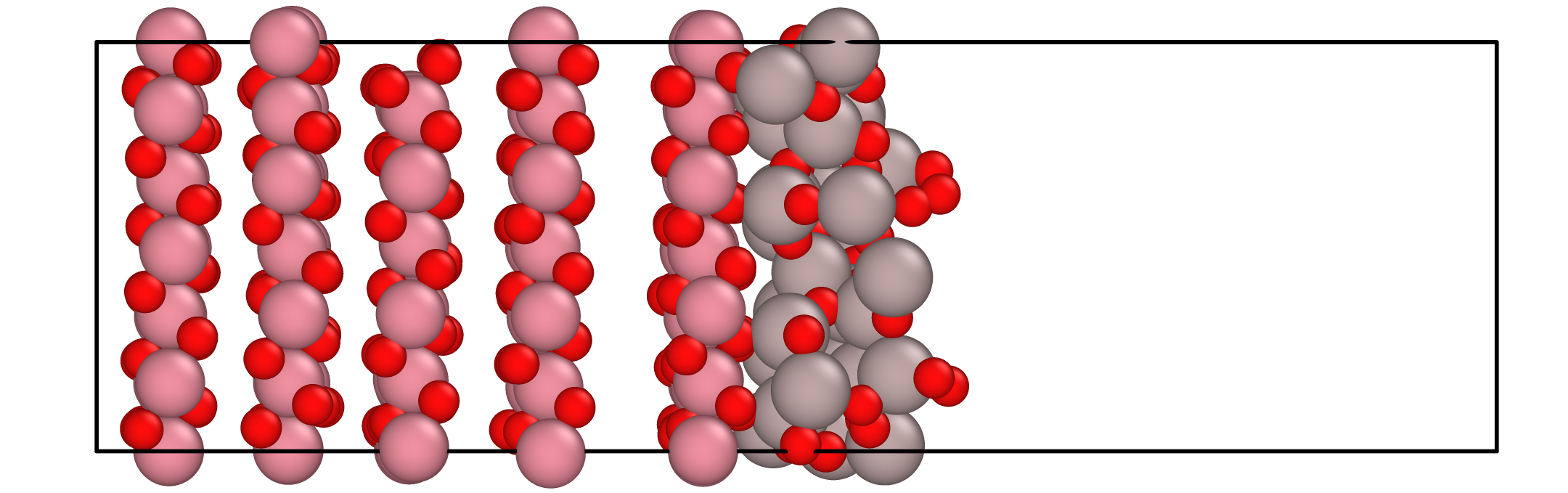}}
\subfloat[Delithiated $\Delta x$~=~4.0~\AA]{\label{MCBJ_F}\includegraphics[width=0.33\columnwidth]{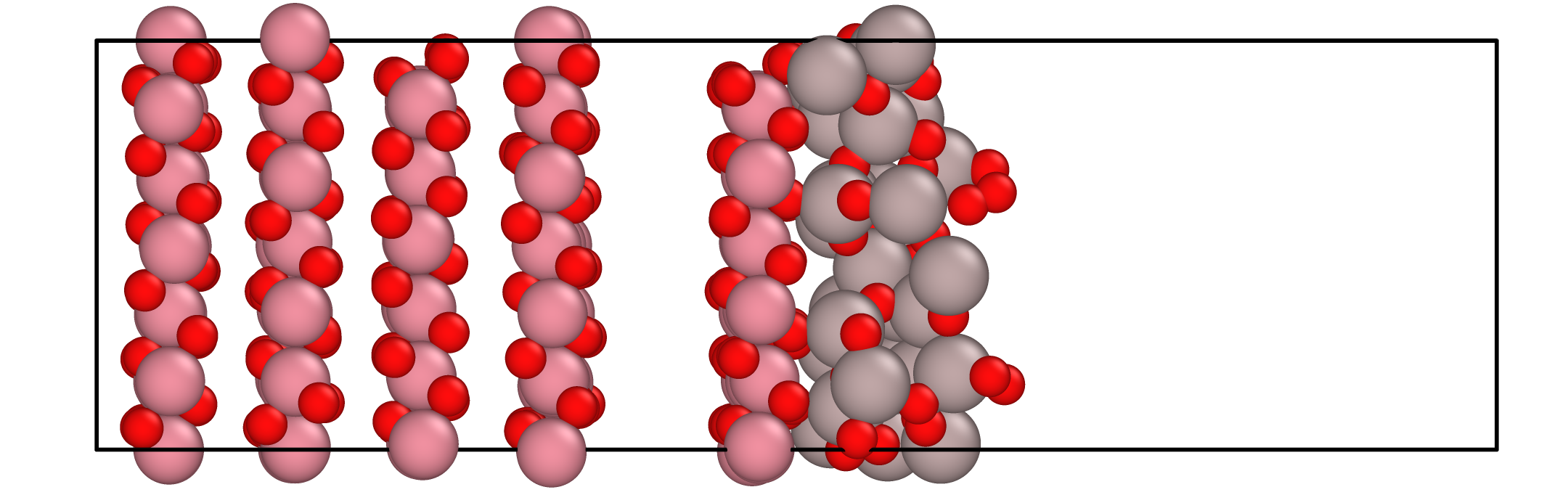}}
\subfloat[Delithiated $\Delta x$~=~8.0~\AA]{\label{MCBJ_G}\includegraphics[width=0.33\columnwidth]{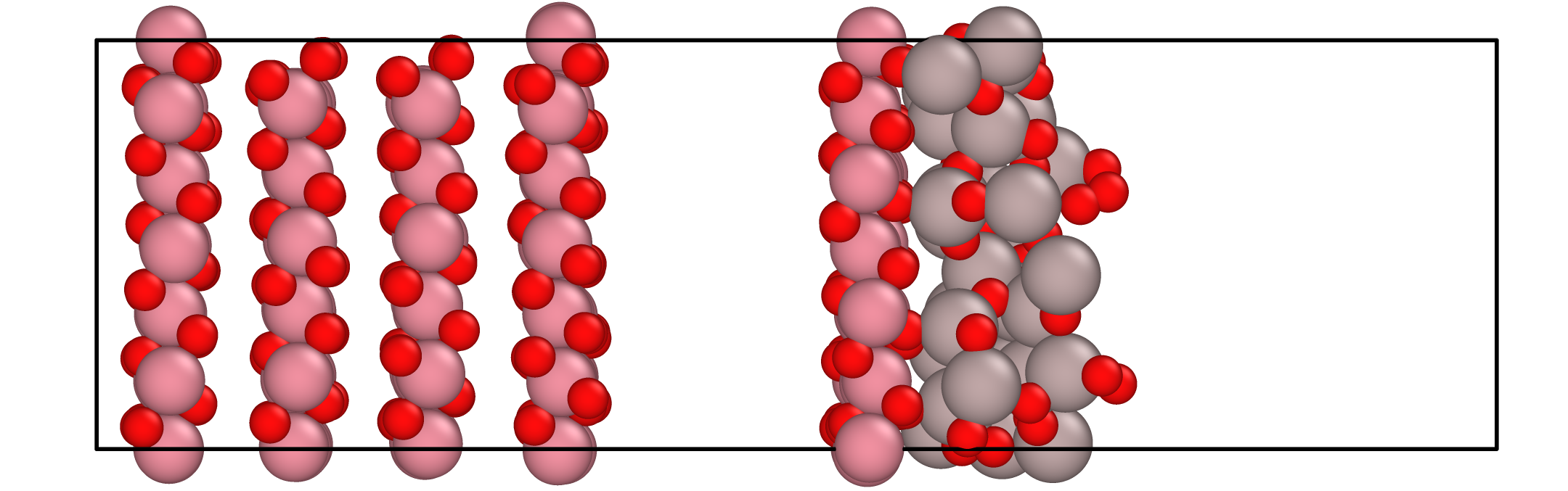}}
\caption{(a) The change in energy ($\Delta E$) as a function of distance stretched ($\Delta x$) for the fully lithiated (black) and fully delithiated (blue) structures, where $\Delta E$ is the difference in energy from the relaxation energy of the system. (b-g) Snapshots of the lithiated and delithiated structures are shown at $\Delta x$ values of 2.0~\AA, 4.0~\AA~and 8.0~\AA.
\label{MCBJ_Fig}}
\end{figure*}

To explore the effect of charge transfer on the lithiation process, we performed charge analysis in each of the cases investigated above. The variation in the net atomic charge of Co atoms in the layer immediately below the coating (layer 5) for each approach is shown in Figure~\ref{Layer5 charge}.
A similarity is observed in the trends for net atomic charge of Co in the ``Alternating" and ``Bottom" approaches. Here, the net atomic charge remains primarily unchanged. However, when the Li atoms closest to the TM atoms located near the coating are drawn out, it seems to affect it more consistently by drawing more electron charge from the Co atoms, showing a gradual increase in the charge transfer of the Co atoms with decreasing Li content. This is the inverse of what we observed for $W$\textsubscript{sep} with the Li content (Figure~\ref{CE_Li}), indicating that the coating adheres more when the interface is more lithiated. More recently, it has been suggested that the composition of the interfacial atomic layers may play a pivotal role in determining the adhesivity of alumina-based Al$_2$O$_3$ coatings in metal alloys \cite{schoeppner_interfacial_2020}.

The results of the MCBJ simulation for the lithiated and delithiated cathodes are shown in Figure~\ref{MCBJ_Fig}.
We find that there are distinct differences in the behavior of the two cathodes based on observations of the energy changes ($\Delta E$) as a function of distance stretched ($\Delta x$) in Figure~\ref{MCBJ_Energy}. For the lithiated cathode, there is not much change in energy through the first 1~\AA~of stretching, followed by a significant energy difference between 1 and 3~\AA. In this region, the topmost layer of the cathode and Li interact very strongly with the Al$_2$O$_3$ layer as it is being removed. This significant change in energy represents all of those interactions and the forces required to fracture the interface. Substantial fluctuations in $\Delta E$ continue through $\approx$~4.5\AA. It is at this point that the Al$_2$O$_3$ coating has fully separated from the cathode. This is why the energy remains relatively constant throughout the simulation. By visualizing the AIMD trajectory, we obtain consistent results, where the Al$_2$O$_3$ layer has begun to move away from the cathode, as shown in Figure~\ref{MCBJ_B}. Even though it is a full 4.0~\AA~ from the surface in Figure~\ref{MCBJ_C}, the Al$_2$O$_3$ layer has reoriented to keep in contact with the cathode surface, but by 8.0~\AA~of separation in Figure~\ref{MCBJ_D}, the Al$_2$O$_3$ has been successfully removed.

The energy profile for the delithiated structure is relatively constant across the entire $\Delta x$ range, indicating that there are no significant changes in atomic interactions as the Al$_2$O$_3$ coating is cleaved.
This is further evident from the snapshots of the AIMD trajectory shown in Figs.~\ref{MCBJ_E} to \ref{MCBJ_G}.
In Figure~\ref{MCBJ_E}, the Al$_2$O$_3$ layer has been moved by 2~\AA, but the top layer of the cathode is pulled with it. This continues for 4.0~\AA~(Figure~\ref{MCBJ_F}) and 8.0~\AA~(Figure~\ref{MCBJ_G}) of stretching, where the cathode layer has been completely separated from the bulk. The behavior of the lithiated and delithiated structures in the MCBJ simulations agrees with the expected outcome determined by calculations of $W$\textsubscript{sep} discussed above.

Overall, our AIMD and cleavage energy calculations show that stripping Li$^+$ from LiCoO$_2$ collapses the inter-layer attraction, mirroring in-situ XRD evidence that above 4.5~V drives the O3→H1-3 phase transition, heavy lattice strain, and eventual particle cracking \cite{wang_enhancing_2024}. The simulations also reproduce the surge of lattice-oxygen release and attendant electrolyte attack seen experimentally, which are key triggers of rapid capacity fade. We find that an ultrathin ALD Al$_2$O$_3$ coating suppresses both oxygen evolution and electrolyte decomposition, thereby preventing the high-voltage phase transition and sustaining structural integrity and capacity over cycling, as reported in recent coated-cathode studies \cite{wang_enhancing_2024,odonoghue_atomic_2023}.

Despite the strong qualitative agreements, we acknowledge the model’s idealized nature and the limitations of direct translation to cell performance. Our simulations probed “worst-case” conditions, for example, a fully delithiated LCO surface, which are beyond what is typically reached in battery operation. In practice, completely removing Li from LCO (charging to 100~\% delithiation) would rapidly trigger catastrophic failure; therefore, cells are rarely pushed to this limit. Nevertheless, the trends observed in the model foreshadow real degradation pathways: the loss of interlayer attraction and layer slippage at 0~\% Li correspond to the gradual c-axis contraction and phase changes seen when LiCoO$_2$  is cycled to high states of charge (e.g., the O3→H1-3 transition)\cite{odonoghue_atomic_2023,chen_advanced_2024}. Over many cycles, these small irreversible changes accumulate as lattice disorder and micro-cracking, much like the single-cycle “shock” of full delithiation in the simulation presages. Likewise, our short-timescale AIMD runs cannot capture slow processes, such as long-term transition-metal migration or progressive crack growth. However, the early signs of Co displacement in the fully delithiated structure point toward the same end states that prolonged cycling yields. We note the limitation of the model's absence of cyclic loading and multi-particle interactions. For example, we observed a reduction in interfacial binding at full delithiation (suggesting a tendency for coating delamination or loss of contact), but only extended cycling experiments can capture the eventual propagation of cracks or detachment of coatings.

Finally, the implications of these findings extend beyond Li-ion batteries. The concept of utilizing ultra-thin, robust coatings to stabilize interfaces is already being explored in other material interfaces and oxide contexts, including thin-film microbattery cathodes, solid-state cells, photoelectrodes, and perovskite solar cells \cite{basnet_asymmetric_2023,feng_mitigating_2025,chen_advanced_2024}. These cross-disciplinary successes underscore that ultra-thin conformal coatings offer a transferable chemical and mechanical passivation paradigm for a wide variety of oxide interfaces.

\subsection{Conclusion}
The application of aluminum oxide coatings on LiCoO$_2$ cathodes has demonstrated significant improvements in both chemical stability and mechanical robustness, which are critical for advancing the performance of lithium-ion batteries. Through AIMD simulations, this study provides atomistic insights into the interaction between organic electrolytes and both coated and uncoated cathode surfaces. The results show that the Al$_2$O$_3$ coating effectively mitigates the degradation of electrolyte molecules, particularly EC and DMC, at the cathode surface. By suppressing electrolyte decomposition and oxygen evolution, the alumina coating stabilizes the cathode-electrolyte interface, particularly in high-charge states, thereby enhancing the cycle life and safety of LIBs.

Mechanically, the alumina coating reinforces the cathode, reducing the tendency for lattice breakdown and crack propagation under cycling stress. The cleavage energy calculations and MCBJ simulations indicate that the coated cathodes exhibit higher interfacial strength and better resistance to mechanical failure compared to uncoated surfaces. The role of coating thickness and the structural orientation of the cathode were found to be critical in optimizing both chemical stability and mechanical durability.

These findings highlight the protective role of Al$_2$O$_3$ coatings in minimizing degradation and enhancing structural integrity. Future work can further explore the precise tuning of coating properties to maximize performance across different cathode materials and cycling conditions, pushing the boundaries of battery technology for next-generation energy storage solutions.

\begin{acknowledgement}

This work was supported by the National Science Foundation under Grant No. CBET-2028722. This work was performed using \textit{Expanse}, a part of the Extreme Science and Engineering Discovery Environment (XSEDE/ACCESS), which is supported by NSF Grant No. 1928224 under allocation TG-DMR180009 and \textit{Spiedie} HPC at Binghamton University.

Note: Certain equipment, instruments, software, or materials are identified in this paper in order to specify the experimental procedure adequately. Such identification is not intended to imply recommendation or endorsement of any product or service by NIST, nor is it intended to imply that the materials or equipment identified are necessarily the best available for the purpose.

\end{acknowledgement}

%
%

\bibliography{References.bib}

\end{document}